\documentclass[12pt]{article}
\usepackage{epsf}
\begin{document}
\newcommand {\sheptitle}
{\bf\large{Validity of Quasi-Degenerate Neutrino Mass Models and their Predictions on Baryogenesis}}
\newcommand {\shepauthor}
{ Ng.K.Francis$^{a,b}$\footnote{\it{E-mail:} ngkf2010@gmail.com}
 and N. Nimai Singh$^{b}$\footnote{
  \it{E-mail:} nimai03@yahoo.com}}

\newcommand{\shepaddress}
{$^{a}$Department of Physics, Tezpur University, Tezpur-784028,India \\
 $^{b}$Department of Physics, Gauhati University,Guwahati-781014,India}
\newcommand{\shepabstract}
 {Quasi-degenerate neutrino mass models (QDN) which can
explain the current data on neutrino masses and mixings, are
studied. In the first part, we study the  effect of CP-phases on
QDN mass matrix ($m_{LL}$) obeying $\mu-\tau$ symmetry in normal hierarchical
(QD-NH) and inverted hierarchical (QD-IH) patterns. The numerical
predictions are consistent with observed data on (i) solar mixing
angle $(\theta_{12})$ which lies below tri-bimaximal (TBM) value, (ii)
absolute neutrino masses consistent with $0\nu\beta\beta$ decay mass
parameter $(m_{ee})$ and (iii) cosmological upper bound  $\sum^{3}_i
m_{i}$.  $m_{LL}$ is parameterized using only two unknown parameters $(\epsilon,\eta)$ within $\mu-\tau$ symmetry. 
The second part deals with the estimation of observed baryon asymmetry
of the universe (BAU) where we consider the Majorana CP violating
phases ($\alpha,\beta$) and the Dirac neutrino mass matrix
$(m_{LR})$. $m_{LR}$  is taken as either the charged lepton or the  up
quark mass matrix. $\alpha,\beta$ is derived from the heavy
right-handed Majorana mass matrix $M_{RR}$. $M_{RR}$ is generated from
$m_{LL}$ and $m_{LR}$ through inversion of Type-I seesaw formula. The
predictions for BAU are nearly 
consistent with observations for flavoured thermal leptogenesis
scenario for Type-IA in both QD-NH and QD-IH models. We also observe
some enhancement effects in flavour leptogenesis compared to
non-flavour leptogenesis by a magnitude of order one. In non-thermal
leptogenesis QD-NH Type-IA is the only model consistent with
observed data on baryon asymmetry. QD-NH model
appears to be more favourable than those of QD-IH. The predicted
inflaton mass needed to produce the BAU is found to be $M_{\phi}\sim
10^{10}$ GeV corresponding to the reheating temperature $T_{R}=
10^{6}$ GeV.  The present analysis shows that the three absolute
neutrino masses may exhibit quasi-degenerate pattern in nature.

%\vspace{0.002in}

{\bf Keywords:} QDN models, absolute neutrino masses, leptogenesis.\\
{\bf PACS Numbers:} 14.60.Pq; 12.15. Ff; 13.40.Em.}

%77777777777777777777777777777777777777777777777777777777777777777777777
\begin{titlepage}
%\begin{flushright}
%\end{flushright}
\begin{center}
{\large{\bf\sheptitle}}
\bigskip \\
\shepauthor
\\
\mbox{}
{\it \shepaddress}
\\
\vspace{.05in}
{\bf Abstract} \bigskip \end{center}\setcounter{page}{0}
\shepabstract
\end{titlepage}

%%%%%%%%%%%%%%%%%%%%%%%%%%%%%%%%%%%%%%%%%%%%%%%%%%%%%%%%%%%%%%%%%%%
%\twocolumn
\section{Introduction}
{The presently available tightest cosmological upper bound of the sum of three
absolute neutrino masses, has come down to the lowest value,
$\sum_im_{i}\leq 0.28$ eV [1], and a larger value of neutrino mass
$m_{i}\geq 0.1$ eV in Quasi-Degenerate Neutrino (QDN) mass models,
 has therefore to be ruled out. Furthermore, the upper bound value of neutrino
 mass parameter $|m_{ee}|\leq 0.27$ eV appeared in the neutrinoless
 double beta decay $(0\nu\beta\beta)$ experiments [2], also disfavours
 larger values of neutrino mass eigenvalues with same CP-parity.
 Investigations on QDN models in normal hierarchical (QD-NH) and
 inverted hierarchical (QD-IH) patterns of the three absolute neutrino
 masses, require a detailed numerical 
 analysis to check whether such QDN models can really accommodate
 lower values of absolute neutrino masses $|m_{i}|\leq 0.09$ eV  which are
 consistent with the above cosmological bound [1]. In the next step, the
 QDN models are again applied for the prediction of baryon asymmetry
 $(\eta_{B})$ of the universe [3]. In order to estimate the observed
 baryon asymmmetry $\eta_{B}=(6.5^{+0.4}_{-0.5})10^{-10}$ [3] from a
 given neutrino mass model, one usually starts with a suitable
 texture of light Majorana neutrino mass matrix $(m_{LL})$ and then
 relates it with the heavy Majorana neutrinos matrix $(M_{RR})$ and
 the Dirac neutrino mass matrix $(m_{LR})$ through the inversion of
 Type-I seesaw mechanism in an elegant way. Since the stucture of
 Yukawa matrix for Dirac neutrino is not known, we consider the
 texture of Dirac neutrino mass matrix $m_{LR}$ as either the
 charged lepton mass matrix or up quark mass matrix, as allowed by
 SO(10) GUT models for phenomenological analysis.
\par

In many of the theoretical calculations on leptogenesis, the complex
CP violating phases are usually derived from the Majorana phases
appearing in PMNS leptonic mixing matrix $U_{PMNS}$ which diagonalizes
$m_{LL}$. Hence in such approach $m_{LL}$ is no longer taken as the
starting point. However, in the present work, we consider a different
route for the origin of complex CP violating phases which are  derived
from $M_{RR}$ in the estimation of baryon asymmetry of the
universe. This  theoretical possibility is the main
part of the present investigations.

\par
We first introduce a general classification of QDN models based on
their CP-parity patterns in the three absolute neutrino masses, and
then parameterize the mass matrices using only two unknown parameters
$(\epsilon,\eta)$, which can reproduce correct predictions on neutrino
oscillation mass parameters and mixing angles, consistent with the
latest observational data. Though such parameterization is intuitive,
it is quite realistic for phenomenological analysis. The estimations
of baryon asymmetry of the universe in the light of thermal and
non-thermal leptogenesis, may thus serve as an additional criteria to
discriminate the correct pattern of neutrino mass models and also to
shed light on the structure of unknown Dirac neutrino mass matrix.
\par
The paper is organised as follows. In section 2 we parameterize the
neutrino mass matrix. In section 3, numerical analysis and predictions
on baryon asymmetry are outlined. Finally in section 4 we conclude
with a summary. In the Appendix we briefly summarize the formalism for
estimating the lepton asymmetry in thermal leptogenesis through
out-of-equilibrium decay of the heavy right-handed Majorana
neutrinos. This is followed by flavoured thermal leptogenesis and non-thermal leptogenesis.

\section{\bf Parameterization and computations}
\subsection{\bf Parameterizations of neutrino mass matrix}

A general $\mu$-$\tau$ symmetric neutrino mass matrix [4,5] with its
four unknown independent matrix elements, requires at least four
independent equations for realistic numerical solution,
\begin{equation}
m_{LL}=
\left(\begin{array}{ccc}
 m_{11}& m_{12} & m_{12}\\
 m_{12} & m_{22} & m_{23}\\
 m_{12} & m_{23} & m_{22}
\end{array}\right).
\end{equation}
The three mass eigenvalues $m_{i}$ and solar mixing angle $\theta_{12}$,
are given by \\
$ m_{1} = m_{11}-\sqrt{2}\tan\theta_{12}m_{12},\hspace{.1in}\\
 m_{2}=m_{11}+\sqrt{2}\cot\theta_{12}m_{12}, \hspace{.1in}\\ 
 m_{3}=m_{22}-m_{23}$.
\begin{equation}
\tan2\theta_{12}=\frac{2\sqrt{2}m_{12}}{m_{11}-m_{22}-m_{23}}.
\end{equation}
The observed mass-squared differences are calculated as\\ 
\begin{equation}
\bigtriangleup m^{2}_{12} = m^{2}_{2}-m^{2}_{1} >0, \hspace{.1in}
\bigtriangleup m^{2}_{32} =\left| m^{2}_{3}-m^{2}_{2}\right|.\\
\end{equation}\\ 
In the basis where charged lepton mass matrix is diagonal, we have the
leptonic mixing matrix, $U_{PMNS} = U$,  where 
\begin{equation}
U_{PMNS} =
 \left(\begin{array}{ccc}
 \cos\theta_{12} & -\sin\theta_{12} & 0\\
\frac{\sin\theta_{12}}{\sqrt{2}} & \frac{\cos\theta_{12}}{\sqrt{2}} & 
-\frac{1}{\sqrt{2}}\\
\frac{\sin\theta_{12}}{\sqrt{2}} & \frac{\cos\theta_{12}}{\sqrt{2}} &
\frac{1}{\sqrt{2}}
\end{array}\right).
\end{equation}     
The neutrino mass parameter $m_{ee}$ in $0\nu\beta\beta$ decay and the sum of the
absolute neutrino masses in WMAP cosmological bound $\sum_i m_{i}$, are
given respectively by,
\begin{equation} 
m_{ee} = 
|m_{1}U^{2}_{e1} + m_{2}U^2_{e2}+ m_{3}U^2_{e3}|, \\\\\\  m_{cosmos}=
m_{1}+m_{2}+m_{3}.
\end{equation}
A general classification for three-fold quasi-degenerate neutrino mass
models [5] with respect to Majorana CP-phases in their three mass
eigenvalues, is adopted here. Diagonalization of left-handed Majorana
neutrino mass matrix $m_{LL}$ in eq.(1) is given by 
$m_{LL} = UDU^{T}$, where U is the diagonalising matrix in eq.(4) and
Diag=D$(m_{1},m_{2}e^{i\alpha}, m_{3}e^{i\beta})$ is the diagonal matrix
with two unknown Majorana phases $(\alpha, \beta)$. In the basis where
charged lepton mass matrix is diagonal, the leptonic mixing matrix is
given by $U= U_{PMNS}$ [6]. We then adopt the following classification
according to their CP-parity patterns in the mass eigenvalues
$m_{i}$ namely Type IA: (+-+) for D=Diag$(m_{1},-m_{2},m_{3})$; Type
IB:(+++) for D=Diag$(m_{1},m_{2},m_{3})$ and Type-IC: for (++-) for
D=Diag$(m_{1},m_{2},-m_{3})$ respectively. We now introduce the
following parameterization for $\mu$-$\tau$ symmetric neutrino mass 
matrices $m_{LL}$ which  satisfy the above classifications [4,5] and
present a detailed numerical analysis.

\subsection {Numerical analysis and predictions}

For detailed numerical analysis we first choose the light Majorana
neutrino mass matrix $m_{LL}$ presented in
section 2.1. These mass matrices which obey $\mu-\tau$ symmetry [4],
have the ability to deviate the solar mixing angle from that of TBM
[7]. Next we estimate the numerical values of the three absolute
neutrino masses. As discussed before, we need to introduce the
neutrino mass scale $m_{3}$ as input parameter in addition to the
observed data [8] on solar and atmospheric neutrino mass-squared
differences $(\bigtriangleup m^{2}_{21}$ and $|\bigtriangleup
m^{2}_{32}|)$. The three input parameters are fixed at the stage of
predictions of neutrino mass parameters and mixing angles given in Tables 2
and 3. These results are consistent with the recent data on neutrino
oscillation parameters [8]. For numerical analysis we use the best-fit
values of the global neutrino oscillation observational data [8] on
solar and atmospheric neutrino mass-squared differences. 
 $\bigtriangleup m^{2}_{12}=\left(m^{2}_{2}
- m^{2}_{1} \right)=7.60 \times 10^{-5} eV^{2}$ and
$\left|\bigtriangleup m^{2}_{32}\right|=\left|m^{2}_{3}-m^{2}_{2}
\right|=2.40\times10^{-3}eV^{2}$, and define the following parameters
 $\rho=\frac{\left|\bigtriangleup m^{2}_{23}\right|}{m^{2}_{3}}$ and
$\psi=\frac{\bigtriangleup m^{2}_{21}}{\left|\bigtriangleup m^{2}_{23}\right|}$,
where $m_{3}$ is the input quantity allowed by the latest cosmological
bound. For QDN in normal hierachy (QD-NH) pattern, the other two mass
eigenvalues are estimated from, 
\begin{equation}
m_{2}=m_{3}\sqrt{1-\rho}; m_{1}=m_{3}\sqrt{1-\rho(1+\psi)}
\end{equation}  and for QDN in inverted hierarchy(IH-QD) the mass
eigenvalues are extracted from,
\begin{equation}
 m_{2}=m_{3}\sqrt{1+\rho}; m_{1}=m_{3}\sqrt{1+\rho(1-\psi)}. 
\end{equation}
For suitable input value of $m_{3}$, one can estimate the numerical
values of $m_{1}$ and $m_{2}$ for both QD-NH and QD-IH cases, using
the observational values of $|\bigtriangleup m^{2}_{23}|$ and  
$\bigtriangleup m^{2}_{21}$. Table-1 gives the calculated numerical
values for both models [9].
%%%%%%%%%%%%%%%%%%%%%%%%%%% TABLE 1: DATA %%%%%%%%%%%%%%%%
\begin{table}
\begin{tabular}{cccccc}  \hline
%\multicolumn{4}{|c|}{Table-1:The masses}.\hline\hline
{input} & {calculated} & 
\multicolumn{2}{c}{QD-NH} &  
\multicolumn{2}{c}{QD-IH}  \\ \cline{3-6}
\bf$m_{3}$ & \bf $\rho$ & \bf $m_{1}$ & \bf $m_{2}$ & 
\bf $m_{1}$ &\bf $m_{2}$ \\  \hline\hline
0.40 & 0.015 & 0.39689 & 0.39699 & 0.40289 & 0.40299\\
0.10 & 0.24 & 0.08674 & 0.08718 & 0.11104 & 0.11136\\
0.08 & 0.375 & 0.06264 & 0.06326 & 0.09340 & 0.09381\\\hline
\end{tabular}
\caption{The absolute neutrino masses  in eV, estimated from
  oscillation data, using calculated $\psi=0.031667$ as defined in the
  text.}
\end{table}
%%%%%%%%%%%%%%%%%%%%%%%%%%%%%%%%

\par  In the next step we parameterize the mass matrix in
eq.(1) into three different types [5,10]: \\
{\bf Type IA}  with D=Diag($m_{1},-m_{2},m_{3}$).
The mass matrix of this type can be parameterized using two
parameters  $(\epsilon,\eta)$:
\begin{equation}
m_{LL}=
\left(\begin{array}{ccc}
\epsilon-2\eta & -c\epsilon & -c\epsilon\\
-c\epsilon &\frac{1}{2}-d\eta & -\frac{1}{2}-\eta\\ 
-c\epsilon & -\frac{1}{2}-\eta& \frac{1}{2}-d\eta 
\end{array}\right)m_{3}.
\end{equation} 
This predicts the solar mixing angle,
\begin{equation}
\tan 2\theta_{12}=-\frac{2c\sqrt{2}}{1+(d-1)\frac{\eta}{\epsilon}}.
\end{equation}
%%%%%%%%%%%%%%%%%%%%%% TABLE 2: Tan= 0.50  %%%%%%%%%%%%%%%%%%%%%
%TABLE-2
\begin{table}
\begin{tabular}{lllll}  \hline
{Different} & 
\multicolumn{2}{c}{QD-NH} &  
\multicolumn{2}{c}{QD-IH}  \\ \cline{2-5}
\bf{parameters} & \bf Type-IA & \bf Type-IB & \bf Type-IA & \bf
Type-IB \\ \hline \hline
c & 1.0 & 1.0 & 1.0 & 1.0\\ 
d & 1.0 & 1.0 & 1.0 & 1.0\\
$m_{3}$ & 0.10 & 0.10 & 0.08 & 0.08\\
$\epsilon$ & 0.57972 & 0.0015 & 0.78004 & 0.00169\\
$\eta$ & 0.14602 & 0.0649 & 0.19628 & -0.08546\\ \hline

$m_{1}$ (eV) & 0.08674 & 0.08675 & 0.09340 & 0.09340\\
$m_{2}$ (eV) & -0.08717 & 0.08717 & -0.09380 & 0.09380\\
$m_{3}(eV)$ & 0.10 &0.10 & 0.08 & 0.08 \\ 
$\sum|m_{i}|eV$ & 0.27 & 0.274 & 0.267 & 0.274\\ \hline
$\bigtriangleup m^{2}_{21}eV^{2}$ & $7.6\times10^{-5}$ &
$7.6\times10^{-5}$ & $7.6\times10^{-5}$ & $7.6\times10^{-5}$\\
$\left|\bigtriangleup m^{2}_{23} \right| eV^{2}$ & $2.2\times10^{-3}$ &
$2.4\times10^{-3}$ & $2.4\times10^{-3}$ & $2.4\times10^{-3}$\\  
$\tan^{2}\theta_{12}$ & 0.50 & 0.50 & 0.50 & 0.50\\
$|m_{ee}|$ eV  & 0.08688 & 0.0869 & 0.09354 & 0.09354\\ \hline
\end{tabular}
\caption{Predictions for $\tan^{2}\theta_{12}=0.50$ and other parameters
  consistent with observations.}
\end{table}

%%%%%%%%%%%%%%%%
When we choose the constant parameters $c=d=1.0$, we set for the
tri-bimaximal mixings (TBM) $tan2\theta_{12}=-2\sqrt{2}$
(which is $\tan^{2}\theta_{12}= 0.50$) and the values of $\epsilon$ and $\eta$
are calculated for both QD-NH and QD-IH models, by using the values 
of obsrvational data [8] in these two eigenvlue expressions:
$m_{1}=(2\epsilon-2\eta)m_{3}$ and $m_{2}=(-\epsilon-2\eta)m_{3}$
obtained for TBM solution in eq.(8).
The results are given in Table-2 for $tan^{2}\theta_{12}=0.50$. 
The solar mixing angle can be further lowered by taking the values $c<1$ 
and $d>1$, while retaining the earlier values of $\epsilon$ and $\eta$ 
extracted for TBM solution.  For $\tan^{2}\theta_{12}=0.45$, case the
results  are shown in Table-3. \\{\bf Type-IB} with 
D = Diag $(m_{1}, m_{2}, m_{3})$: This class of
quasi-degenerate mass pattern is given by the mass matrix, 
\begin{equation}
m_{LL} = \left(\begin{array}{ccc}
1-\epsilon-2\eta & c\epsilon & c\epsilon\\ 
c\epsilon &  1-d\eta & -\eta\\ 
c\epsilon & -\eta & 1-d\eta
\end{array} \right)m_{3}. 
\end{equation}  
This predicts the solar mixing angle, 
\begin{equation}
\tan2\theta_{12} =
\frac{2c\sqrt{2}}{1+(1-d)\frac{\eta}{\epsilon}}.                
\end{equation}
which gives the TBM solar mixing angle with the input values c = 1 and
d = 1. When $\epsilon=0$, $\eta=0$, this leads to
$m^{diag}_{LL}=diag(1,1,1)m_{3}$.
Like in Type-IA, here {\bf $\epsilon$} and {\bf $\eta$}
values are computed for QD-NH and QD-IH, by using observational
data [8] in the corresponding two eigenvalue expressions: 
$m_{1}=(1-2\epsilon-2\eta)m_{3}$ and $m_{2}=(1+\epsilon-2\eta)m_{3}$
for TBM solution in eq.(10).

%%%%%%%%%%%%%%%%%%%%%%% TABLE 3: TAN = 0.45 %%%%%%%%%%%%%%%%
\begin{table}
\begin{tabular}{lllll}  \hline
{Different} & 
\multicolumn{2}{c} {QD-NH} &  
\multicolumn{2}{c} {QD-IH}  \\ \cline{2-5}
\bf {parameters} & \bf Type-IA & \bf Type-IB & \bf Type-IA &
 \bf Type-IB \\ \hline\hline 

c & 0.868 & 0.945 & 0.868 & 0.96\\
d & 1.025 & 0.998 & 1.0 & 1.002\\
$m_{3}$ & 0.10 & 0.10 & 0.08 & 0.08\\
$\epsilon$ & 0.6616 & 0.00145 & 0.88762 & 0.00169\\
$\eta$ & 0.1655 & 0.06483 & 0.22317 & -0.08546\\ \hline

$m_{1}$ (eV) & 0.0876 & 0.08676 & 0.09392 & 0.09341 \\
$m_{2}$ (eV) & -0.0880 & 0.08717 & -0.09432 & 0.09381\\
$m_{3}(eV)$ & 0.0996 &0.10002 & 0.08 & 0.080014 \\ 
$\sum|m_{i}|eV$ & 0.274 & 0.274 & 0.268 & 0.267\\ \hline
$\bigtriangleup m^{2}_{21}eV^{2}$ & $7.7\times10^{-5}$ &
$7.3\times10^{-5}$ & $7.6\times10^{-5}$ & $7.4\times10^{-5}$\\
$\left|\bigtriangleup m^{2}_{23} \right| eV^{2}$ & $2.2\times10^{-3}$ &
$2.4\times10^{-3}$ & $2.4\times10^{-3}$ & $2.4\times10^{-3}$\\  
$\tan^{2}\theta_{12}$ & 0.45 & 0.45 & 0.45 & 0.45\\
$|m_{ee}|$ eV  & 0.0877 & 0.08688 & 0.09403 & 0.09354\\ \hline
\end{tabular}
\caption{Predictions for $\tan^{2}\theta_{12}=0.45$ and other parameters
  consistent with observations.}
\end{table}
%\newpage
%%%%%%%%%%%%%%%%%%%%% END TABLE 3 %%%%%%%%%%%%%%%%%%%%%%%%%%%%%%%%%%

{\bf Type-IC} with D = Diag($m_{1},m_{2},-m_{3}$): It is not
necessary to treat this model separately as it is similar to
Type-IB except with the interchange of two matrix elements ($m_{22}$)
and $(m_{23})$ in the mass matrix in eq.(10), and this effectively
imparts an additional odd CP-parity on the third mass eigenvalue
$m_{3}$ in Type-IC. Such change does not alter the predictions of
Type-IB. Tables 2 and 3 present our numerical results for  
$tan^{2}\theta_{12}=0.50$ and $tan^{2}\theta_{12}=0.45$ respectively which are
consistent with cosmological bound.

%%%%%%%%%%%%%%%%%%%%%%%%%%%%%%%%%%%%%%%%%%%%%%%%%%%%%%%%%%%%%%%%%%%%%%

%%%%%%%%%%%%%%%%%%%%%%%%% TABLE 4: Majorana Mass Mi %%%%%%%%%%%%%%%%%%
\begin{table}
\begin{tabular}{ccccl} \hline
{Type} & (m,n) &  $M_{1}$ (GeV)& $M_{2}$ (GeV) &  $M_{3}$ (GeV) \\ \hline
NH-IA & (6,2) & $ 4.8659\times10^{8}$ & $-3.5068\times10^{12}$ &
 $ 9.1256\times10^{14}$ \\
  &  (8,4) & $3.9414\times10^{6}$ & $-3.9774\times10^{10}$ &
 $6.0097\times10^{13}$  \\ \hline 
NH-IB & (6,2) & $ 1.8117\times10^{8}$ & $2.6219\times10^{12}$ &
 $ 3.2528\times10^{14}$ \\
  &  (8,4) & $1.4994\times10^{6}$ & $2.1238\times10^{10}$ &
 $ 3.2527\times10^{14}$  \\ \hline 
IH-IA & (6,2) & $ 4.5568\times10^{8}$ & $-2.8771\times10^{12}$ & 
$ 1.3273\times10^{14}$ \\
  &  (8,4) & $3.6910\times10^{6}$ & $-2.3241\times10^{10}$ & $
1.8289\times10^{14}$  \\ \hline 
IH-IB & (6,2) & $ 1.7153\times10^{8}$ & $2.8234\times10^{12}$ &
 $3.5098\times10^{14}$ \\
  &  (8,4) & $41.393\times10^{6}$ & $2.4809\times10^{10}$ & $
3.8091\times10^{14}$  \\ \hline 
\end{tabular}
\caption{Heavy right-handed Majorana neutrino mass
  $M_{j}$  for QDN with normal and inverted ordering mode for
  $\tan^{2}\theta_{12}=0.45,$ using neutrino mass matrices given in
  section 2. The entry (m,n) indicates the type of Dirac neutrino mass
  matrix, as explained in the text.}
\end{table}

\section{\bf  Predictions of baryon asymmetry}
We now apply the above Quasi-degenerate  neutrino mass matrices with
the values of the parameters already fixed in section 2, in the calculation of
baryon asymmetry of the universe. This gives the second stage for the
discrimination among the QDN models.

   For the numerical calculation of baryon asymmetry, we refer to  all the
   relevant expressions given in the Appendix. First we translate mass
matrices $m_{LL}$ via the inversion of the seesaw formula [11], 
$M_{RR}=-m_{LR}^{T}m^{-1}_{LL}m_{LR}$. We choose a basis for $U_{R}$ where
$M^{diag}_{RR}= U^{T}_{R}M_{RR}U_{R}=diag(M_{1},M_{2},M_{3})$, with
real and positive mass eigenvalues. We then transform diagonal form of Dirac
neutrino mass matrix, $m_{LR}= diag(\lambda^{m},\lambda^{n},1)\nu$ to the $U_{R}$
basis: $m_{LR}\rightarrow m^{'}_{LR}= m_{LR}U_{R}Q $ where $Q = diag(1,
e^{i\alpha},e^{i\beta})$ is the complex matrix containing
CP-violating Majorana phases introduced by hand. Here $\lambda$
is the Wolfenstein parameter and the choice (m,n) in $m_{LR}$ gives
the type of Dirac neutrino mass matrix. The value of the vev is taken
as $v=174$ GeV.
\par
At the moment we consider phenomenologically two possible forms of Dirac
neutrino mass matrices such as (i) (m,n) = (6,2) for the charged-lepton
type mass matrix, and (ii) (8,4) for up-quark type mass matrix. In this
prime basis the Dirac neutrino Yukuwa coupling becomes $h =
\frac{h^{'}_{LR}}{\nu}$ which enters in the expression of CP-asymmetry
$\epsilon$ in (14) and (23) given in the Appendix. The new Yukawa
coupling matrix h also becomes a complex quantity, and hence the term
$Im(h^{\dag}h)_{1j}$ appearing in lepton asymmetry $\epsilon_{1}$,
gives a non-zero contribution. A straight forward simplification shows that
$(h^{\dag}h)^{2}_{1j}=(Q^{*}_{22})^{2}Q^{2}_{22}R_{2}+(Q^{*}_{11})^{2}Q^{2}_{33}R_{2}$ where $R_{2,3}$ are real parameters. After inserting the values of
phases, the above expression leads to $Im(h^{\dag}h)^{2}_{1j}=
-[R_{2}sin2(\alpha-\beta)+R_{3}sin2\alpha]$ which imparts 
a non-zero CP asymmetry for particular choice of $(\alpha,\beta)$.

%%%%%%%%%%%%% %%%   TABLE 5 Majorana Mass    %%%%%%%%%%%%%%%%%%%%%%
\begin{table}
\begin{tabular}{lllllll} \hline
 \emph{Type} & (m,n) & $(h^{\dag}h)_{11}$  & $\epsilon_{1}$ &
$\eta_{1B}$ & $\eta_{3B}$ \\ \hline
NH-IA & (6,2) & $3.73\times10^{-6}$  & $1.92\times
10^{-7}$ & $9.07\times10^{-12}$ & $2.11\times 10^{-11}$  \\
  &  (8,4) & $3.03\times10^{-8}$  & $1.55\times10^{-9}$ &
$7.32\times10^{-14}$ & $1.71\times10^{-13}$ \\ \hline 
NH-IB & (6,2) & $ 5.31\times10^{-7}$  & $1.12\times
10^{-14}$ & $1.41\times10^{-18}$ & $ 5.67\times 10^{-13}$  \\ %\cline{2-7}
 &  (8,4) & $4.30\times10^{-9}$  & $ 8.87\times10^{-17}$ &
$1.12\times10^{-20}$ & $ 4.71\times10^{-15}$ \\ \hline
IH-IA & (6,2) & $3.77\times10^{-6}$  & $1.94\times
10^{-7}$ & $8.50\times10^{-12}$ & $1.98\times 10^{-11}$  \\ %\cline{2-7}
  & (8,4) & $3.05\times10^{-8}$  & $1.57\times10^{-9}$ &
$6.88\times10^{-14}$ & $1.60\times10^{-13}$ \\ \hline
IH-IB & (6,2) & $5.31\times10^{-7}$  & $ 9.75\times
10^{-15}$ & $ 1.15\times10^{-18}$ & $ 5.95\times 10^{-13}$  \\ %\cline{2-7}
  & (8,4) & $4.30\times10^{-9}$  & $ 7.80\times10^{-17}$ &
$ 9.17\times10^{-21}$ & $ 4.76\times10^{-15}$ \\ \hline
\end{tabular}
\caption {Values of CP asymmetry $\epsilon_{1}$ and baryon asymmetry ($\eta_{1B},\eta_{3B}$) for all quasi-degerate models, with $\tan^{2}\theta_{12}=0.45$, using neutrino mass matrices given in the text.The entry (m,n) in $m_{LR}$
  indicates the type of Dirac neutrino mass matrix taken as charged
  lepton mass matrics (6,2) or up quark mass matrix (8,4) as explain
  in the text. }.
\end{table}

%%%%%%%%%%%%%%%%%%Table 6: Inflaton Mass %%%%%%%%%%%%%%%%%%%%%%%%%%%%%%%%%%%%%%%
\begin{table}
\begin{tabular}{llll} \hline
 \emph{Type} & (m,n) & $ T^{min}_{R}$ $<$ $T^{R}\leq T^{max}_{R}(GeV)$  &
 $M^{min}_{\phi}$ $<$ $M_{\phi}\leq M^{max}_{\phi} (GeV) $ \\ \hline
NH-IA & (6,2) & $ 5.51\times10^{6}$ $<$ $T_{R}\leq 4.87\times10^{7}$ & $
9.73\times10^{9}$ $<$ $ M_{\phi} \leq 8.59\times 10^{10}$ \\
  &  (8,4) & $ 2.21\times10^{2}$ $<$ $T_{R}\leq 3.94 \times10^{4}$ &
$ 7.88\times10^{6}$ $<$ $ M_{\phi} \leq 1.40\times10^{9}$ \\ \hline
NH-IB & (6,2) & $ 3.7\times10^{8}$ $<$ $T_{R}\leq 1.85\times10^{6}$ & $
3.70\times10^{8}$ $<$ $ M_{\phi} \leq 1.89\times 10^{2}$ \\
  &  (8,4) & $ 2.96\times10^{13}$ $<$ $T_{R}\leq 1.49\times10^{4}$ &
$ 2.99\times10^{6}$ $<$ $ M_{\phi} \leq 1.52\times10^{-2}$ \\ \hline
IH-IA & (6,2) & $ 5.12\times10^{5}$ $<$ $T_{R}\leq 4.56\times10^{6}$ & $
9.11\times10^{8}$ $<$ $ M_{\phi} \leq 8.11\times 10^{9}$ \\
  &  (8,4) & $ 5.12\times10^{5}$ $<$  $T_{R}\leq 3.69\times10^{4}$ &
$ 7.38\times10^{6}$ $<$ $ M_{\phi} \leq 5.32\times10^{5}$ \\ \hline
IH-IB & (6,2) & $ 3.84\times10^{12}$ $<$ $T_{R}\leq 1.72\times10^{7}$ & $
3.44\times10^{8}$ $<$ $ M_{\phi} \leq 1.54\times 10^{2}$ \\
  &  (8,4) & $ 3.88\times10^{12}$ $ <$ $T_{R}\leq 1.39\times10^{4}$ &
$ 2.79\times10^{6}$ $< $$ M_{\phi} \leq 9.99\times10^{-3}$ \\ \hline
\end{tabular}
\caption{Theoretical bound on reheating temperature $T_{R}$ and
  inflaton masses $M_{\phi}$ in non-thermal leptogenesis,
  calculated using data from Table 5, for all neutrino mass models with
 $\tan^{2}\theta_{12}=0.45$.}
\end{table}
%%%%%%%%%%%%%%%%%%%%%%%%%%%%%%%%%%%%%%%%%%%%%%%%%%%%%%%%%%%%%%%%%%%%%%%%
 \par
 In our numerical estimation of lepton asymmetry, we chose some
arbitrary values of $\alpha$ and $\beta$ other than
$\frac{\pi}{2}$ and $ 0$. For example, light neutrino masses
$(m_{1},-m_{2},m_{3})$ of Type-IA model, leads to
 $M^{diag}_{RR}=diag(M_{1},-M_{2},M_{3})$, and we thus fix the Majorana phase
 $Q = diag(1,e^{i\alpha}, e^{i\beta}) = diag(1,e^{i(\pi/2+\pi/4)},e^{i\pi/4})$
 for $\alpha = (\pi/2+\pi/4)$ and $\beta = \pi/4$. The extra phase
 $\pi/2$ in $\alpha$ absorb the negative sign before heavy Majorana
 mass. In  our search programmes such a choice of the phase leads to highest
 numerical estimation of lepton CP asymmetry.

 In Table 4 we give numerical prediction on three heavy right-handed
 Majorana neutrino masses from these neutrino mass models under
 consideration for the case of $\tan^{2}\theta_{12}=0.45.$ The three
 heavy right-handed Majorana mass matrices which are constructed
 through the inversion of Type-I seesaw mechanism, for 
two choices of diagonal Dirac-neutrio mass matrix discussed
before. The corresponding baryon asymmetries $\eta_{B}$ are estimated
for both non-flavour $\eta_{1B}$ and flavour $\eta_{3B}$ leptogenesis
respectively in Table 5. As expected, there is an enhancement in baryon
asymmerty by a magnitude of order one (approximately) when
flavour dynamics is included (see Table 5 in Type-IA model with charged lepton
mass matrix) [12,13]. Type-IA with charged lepton Dirac neutrino mass matrix, is
the only model sensitive and consistent with data on observed baryon asymmetry.
\par 

In case of non-thermal leptogenesis, the lightest right-handed Majorana
neutrino mass $M_{1}$ from Table 4 and the CP asymmetry $\epsilon_{1}$
from Table 5, for all the neutrino mass models, are used in the
calculation of the bounds: 
$T^{min}_{R}<T_{R}\leq T^{max}_{R}$ and $M^{min}_{\phi}<M_{\phi}\leq
M^{max}_{\phi}$ in Table 6. The baryon asymmetry 
$Y_{B}=\frac{n_{B}}{s}= 8.7 \times 10^{-11}$ is taken as input value
from WMAP obseervational data. Only those neutrino mass models which
simultaneously satisfy the two constraints, $T^{max}_{R}>T^{min}_{R}$
and $M^{max}_{\phi}>M^{min}_{\phi}$, could survive in the non-thermal
leptogenesis scenario. The predicted inflaton mass as $M_{\phi} \sim
10^{11}$ GeV for reheating temperature $T _{R}=10^{6}$ GeV are  needed
to produce the observed baryon asymmetry of the universe [14,15]. From
Table 6, the neutrino mass nodels with (m,n) which are compatible with
$M_{\phi}\sim(10^{10}-10^{13})$ GeV and $T_{R}\approx(10^{6}-10^{7})$
GeV are listed as NH-IA (6,2) and IH-IA (6,2) only. Again in order to
avoid gravitino problem [8] in supersymmetric models, one has the
bound on reheating temperature, $T_{R}\approx(10^{6}-10^{7})$
GeV. This implies that Type-IA where Dirac
neutrino mass matrix is taken as charged lepton mass matrix is the
only model consistent with the observed baryon
asymmetry. These findings nearly agree with flavoured thermal leptogenesis
for Type NH-IA (6,2) model in Table 5. This result is also consistent
with quasi-degenerate in inverted hierachical (IA) models for charged lepton mass matrix (6,2) [15].

\section{Summary and Conclusions}
We have studied the effects of Majorana CP phases on the
prediction of absolute neutrino masses in three types of QDN models
having both normal and inverted hierarchical patterns within $\mu$-$\tau$
symmetry. These predictions are consistent with data on the mass
squared difference derived from various oscillation experiments, and
from the upper bound on absolute neutrino masses in $0\nu\beta\beta$
decay as well as cosmological upper bound of $\sum_i m_{i}\leq 0.28$ eV. It has been found that QDN models with
$m_{i}\leq 0.09 $ eV, are still far from discrimination and hence the
quasi-degenerate pattern is not yet ruled out. The prediction on solar
mixing angle is also found to be lower than TBM value viz,
$\tan^{2}\theta_{12}$ =0.45 which coincides with the best-fit in the
neutrino oscillation data [8].

 In the next stage, left-handed Majorana neutrino mass matrices
 $m_{LL}$ have been 
 employed to estimate the baryon asymmetry
of the universe, in both thermal and non-thermal leptogenesis scenario
(Tables 5-6). We use the CP violating Majorana phases derived from
right-handed Majorana mass matrix, and also Dirac neutrino mass matrix
as either charged lepton mass matrix or up-quark mass matrix. We also
observe some 
enhancement effects in flavour leptogenesis [20] compared to
non-flavour leptogenesis by a magnitude of order one. The predicted
inflaton mass needed to produce the observed baryon asymmetry of the
universe is found to be $M_{\phi}\sim10^{10}$ GeV corresponding to the
reheating temperature $T_{R} = 10^{6}$ GeV [14,15]. The analysis shows
that quasi-degenerate model in normal hierarchical pattern (NH-IA) 
with charged lepton Dirac mass matrix (6,2),  appears to be a favourable choice
in nature. The quasi-degenerate model in inverted heirarchy(IH-IA)  with the
charged lepton Dirac mass matrix (6,2), is not completely ruled out.
 The results presented in this article are new and have subtle hints
 in the discrimination of neutrino mass models. This could establish
 the quasi-degenerate neutrinos as natural physical neutrinos in the neutrino oscillation experiments.

\vspace{.1in}
{\large \bf Acknowledgement}\\
Ng.K.F thanks the University Grants Commission, Govenment of India for
sanctioning the project entitled 
{\bf ``Neutrino masses and mixing angles in neutrino oscillations''}
vide Grant. No 32-64/2006 (SR). This work was carried out through this
project.

\section*{Appendix}
\subsection*{\bf Thermal leptogenesis}
The light left-handed Majoarana neutrino mass matrix $m_{LL}$ and
heavy right-handed Majorana mass matrix $M_{RR}$ are related through
the canonical seesaw formula (known as Type I) [11] in a simple way:
\begin{equation}
m_{LL}=-m_{LR}M_{RR}m^{T}_{LR}
\end{equation}
where $m_{LR}$ is the Dirac neutrino mass matrix. For our calculation
of lepton asymmetry, we consider the model [16,17] where the asymmetry
decay of the lighest of the heavy right-handed Majorana neutrinos, is
assumed. The physical Majorana neutrino $N_{R}$ decays into two modes:
$N_{R}\rightarrow l_{L}+\bar{\varphi}$, $N_{R}\rightarrow\bar{l}_{L}+\varphi$
where $l_{L}$ is the lepton and $\bar{l}_{L}$ is the antilepton;
$\varphi$ and $\bar{\varphi}$ are the Higgs and anti-Higgs particles respectively. The braching ratio for these two decay
modes are likely to be different. The CP asymmetry which is caused by
the interference of tree level with one-loop corrections for the
decays of the ligh
test of heavy right-handed Majorana neutrino $N_{R}$ is
defined by [16,18]

\begin{equation}
\epsilon=\frac{\Gamma-\bar{\Gamma}}{\Gamma+\bar{\Gamma}}
\end{equation}
Here $\Gamma=\Gamma(N_{i}\rightarrow l_{L}\bar{\varphi})$ and
$\bar{\Gamma}=\Gamma(N_{i}\rightarrow \bar{l}_{L}\varphi)$ are the
decay rates. A perturbative calculation from the interference between
tree level and vertex plus self energy diagrams, gives [19] the lepton
asymmetry $\epsilon_{1}$ for non-SUSY case as
\begin{equation}
\epsilon_{i}=
\frac{1}{8\pi}\frac{1}{(h^{\dag}h)}_{ii} \sum_{j=2,3} Im[(h^{\dag}h)_{y}]^{2}
     [f(\frac{M^{2}_{j}}{M^{2}_{i}}) g(\frac{M^{2}_{j}}{M^{2}_{i}})]
\end{equation}
 where f(x) and g(x) represent the contributions from vertex and self
 energy corrections respectively,$ f(x)
 =\sqrt{x}[-1+(x+1)ln(1+\frac{1}{x})], g(x) = \frac{\sqrt{x}}{x-1}$.
For hierarchical right-handed neutrino masses where x is large, we
have the approximation [16], $f(x)+g(x)\cong
\frac{3}{2\sqrt{x}}$. This simplifies to 
\begin{equation}
\epsilon_{i}=
-\frac{3}{16\pi}[\frac{Im[(h^{\dag}h)^{2}_{12}]}{(h^{\dag}h)_{11}}
\frac{M_{1}}{M_{2}}+\frac{Im[(h^{\dag}h)^{2}_{13}]}{(h^{\dag}h)_{11}}\frac{M_{1}}{M_{3}}]
\end{equation}
where $h=\frac{m_{LR}}{\nu}$ is the Yukawa coupling of the Dirac
neutrino mass matrix in the diagonal basis of $M_{RR}$ and $\nu$= 174
GeV is the vev of the standard model. In term of light Majorana
neutrino mass matrix $m_{LL}$ the above expression can be simplified
to 
\begin{equation}
\epsilon_{1}=-\frac{3}{16\pi}\frac{M_{1}}{(h^{\dag}h)_{11}}
Im[(h^{\dag}m_{LL}h^{*})_{11}]
\end{equation}

For quasi-degenerate spectrum of the heavy right-handed Majorana
neutrino masses, i.e., for $ M_{1}\approx M_{2} < M_{3}$ the asymmetry
is largely enhanced by a resonance factor and in such situation, the
lepton asymmetry is modified [21] to
\begin{equation}
\epsilon_{1}=\frac{1}{8\pi}\frac{Im[(h^{\dag}h)^{2}_{12}]}{(h^{\dag}h)_{11}}
R
\end{equation}
 where
 $R=\frac{M^{2}_{2}(M^{2}_{2}-M^{2}_{1})}{(M^{2}_{1}-M^{2}_{2})^{2}+\Gamma^{2}_{2}M^{2}_{1}}$
 and $\Gamma_{2} =\frac{(h^{\dag}h)_{22}M_{2}}{8\pi}$. It can be noted
 that in case of SUSY, the functions f(x) and g(x) are given by $f(x)=
 \sqrt{x}ln(1+\frac{1}{x})$ and $g(x)=\frac{2\sqrt{x}}{x-1}$; for
 large x one can have $f(x) + g(x)\approx 3/\sqrt{x}$. Therefore the
 factor 3/8 will appear in place of 3/16 in the expression of CP
 asymmetry [20] in eq. (15). The CP asymmetry parameter $\epsilon_{i}$
 is related to leptonic asymmetric parameter $Y_{L}$ as 
\begin{equation}
Y_{L}\approx
\frac{n_{L}-\bar{n}_{L}}{s}=\sum^{3}_{i}\frac{\epsilon_{i}\kappa_{i}}{g_{*i}}
\end{equation}
where $n_{L}$ is the lepton number density, $\bar{n}_{L}$ is the
anti-lepton number density, s is the entropy density, $\kappa_{i}$ is
the dilution factor for the CP asymmetry $\epsilon_{i}$, and $g_{*i}$
is the effective number of degrees of freedom at temperature $T =
M_{i}$. The baryon asymmetry $Y_{B}$ produced through the sphaleron
transition of $Y_{L}$ while the quantum number B-L remains conserved,
is given  by [22]
\begin{equation}
Y_{B}=\frac{n_{B}}{s}= CY_{B-L} = CY_{L}
\end{equation}
where $C = \frac{8N_{F}+4N_{H}}{22 N_{F}+13N_{H}}$, $N_{F}$ is the
number of fermion families and $N_{H}$ is the number of Higgs
doublets. Since $s = 7.04 n_{\gamma}$ the baryon number density over
photon number density $n_{\gamma}$ corresponds to the observed baryon
asymmetry of the universe [23],
\begin{equation}
\eta^{SM}_{B}=(\frac{n_{B}}{n_{\gamma}})^{SM}\approx
d\kappa_{1}\epsilon_{1}
\end{equation}
where $d\approx 0.98\times10^{-2}$ is used in the present
calculation. In case of MSSM, there is no major numerical change with
respect to the non-supersymmetric case in the estimation of baryon
asymmetry. One expects approximate enhancement factor of about
$\sqrt{2}\sqrt{2}$ for strong (weak) washout regime [20].
\par
In the expression for baryon-to-photon ratio in eq. (20), $\kappa_{1}$
describes the washout factor of the lepton asymmetry due to various
lepton number violating processes. This efficiency factor (also known
as dilution factor) mainly depends on the effective neutrino mass

\begin{equation}
\widetilde{m}_{1}=\frac{(h^{*}h)_{11}\nu^{2}}{M_{1}}
\end{equation}

where $\nu$ is the electroweak vev, $\nu$ = 174 GeV. For $10^{-2} eV <
\widetilde{m}_{1} < 10^{3} eV$, the washout factor $\kappa_{1}$ can be well
approximately by [19,24] 

\begin{equation}
\kappa_{1}(\widetilde{m}_{1})=0.3[\frac{10^{-3}}{\widetilde{m}_{1}}][log\frac{\widetilde{m}_{1}}{10^{-3}}]^{-0.6}
\end{equation}
The value of $\kappa_{1}$ is valid only for the given range of $\widetilde{m}$ [25].

%%%%%%%%%%%%%%%%%%%%%%%%%%%TA%B%LE%5%%%%%%%%%%%%%%%%%%%%%%%%
\subsection*{\bf Flavoured thermal leptogenesis}
It is inevitable to include the flavour effects in thermal
leptogenesis [26] and study its effects on the enhancement in baryon
asymmetry over the single flavour approximation. In the flavour
basis the equation for lepton asymmetry in $ N_{1}\rightarrow
l_{\alpha}\varphi$ decay where $\alpha=(e,\mu,\tau)$ becomes
\begin{equation}
\epsilon_{\alpha\alpha}=\frac{1}{8\pi}\frac{1}{(h^{\dag} h)_{11}}
[\sum_{j=2,3}Im[h^{*}_{\alpha 1}(h^{\dag}h)_{1j}h_{\alpha j}]
g(x_{j})+\sum_{j}Im[h^{*}_{\alpha 1}(h^{\dag}h)_{j1}h_{\alpha j}]
\frac{1}{(1-x_{j})}]
\end{equation}
Here we have $x_{j}=\frac{M^{2}_{j}}{M^{2}_{i}}$ and
 $g(x_{j})\sim\frac{3}{2}\frac{1}{\sqrt{x_j}}$.
The efficiency factor for the out-of-equilibrium
situation is given by
 $\kappa_{\alpha}=\frac{m_{*}}{\widetilde{m}_{\alpha\alpha}}$.
Here $\frac{8\pi H\nu^{2}}{M^{2}_{1}}\sim 1.1\times 10^{-3}$ eV, and
  $m_{\alpha\alpha}=\frac{h^{\dag}_{\alpha 1}h_{\alpha 1}}{M_{1}} \nu^{2}$.
This leads to the baryon asymmetry of the universe,
\begin{equation}
\eta_{3B}=\frac{\eta_{B}}{\eta_{\gamma}}\sim
10^{-2}\sum_{\alpha}\epsilon_{\alpha\alpha}\kappa_{\alpha}\sim10^{-2}
m_{*}\sum_{\alpha}\frac{\epsilon_{\alpha\alpha}}{\widetilde{m}_{\alpha\alpha}}
\end{equation}
For single flavour case, the second term in $\epsilon_{\alpha\alpha}$
vanishes when summed over all flavours. Thus
\begin{equation}
\epsilon_{1}\equiv\sum_{\alpha}\epsilon_{\alpha\alpha}=\frac{1}{8\pi}
\frac{1}{(h^{\dag}h)}_{11}\sum_{j}Im[(h^{\dag}h)^{2}_{lj}]g(x_{j}).
\end{equation}
This leads to baryon asymmetry,
\begin{equation}
\eta_{1B}\approx10^{-2}m_{*}\frac{\eta_{1}}{\widetilde{m}}=10^{-2}\kappa_{1}\epsilon_{1}
\end{equation}
where $\epsilon_{1}=\sum_{\alpha} \epsilon_{\alpha\alpha}$ and
$\widetilde{m}=\sum_{\alpha}\widetilde{m}_{\alpha\alpha}$.
\section*{\bf Non-themal leptogenesis}
   We now focus our attention on the non-thermal leptogenesis
   scenario [27] where the right-handed neutrinos are produced through
   the direct non-thermal decay of the inflaton $\phi$. We follow the
   standard procedure [28] where non-thermal leptogenesis and baryon
   asymmetry in the universe had been studied in different neutrino
   mass models whereby some mass models were excluded using bound from
   below and from above on inflaton mass and reheating temperature
   after inflation. We start with the inflaton decay rate given by
\begin{equation}
\Gamma_{\phi}=\Gamma(\phi\rightarrow N_{i}N_{i})\approx
\frac{|\lambda_{i}|^{2}}{4\pi M_{\phi}}
\end{equation}
where $\lambda_{i}$ are the Yukawa coupling constants for the
interaction of three heavy right-handed neutrinos $N_{i}$ with the
inflaton $\phi$ of mass $M_{\phi}$. The reheating temperature after
inflation is given by the expression,
\begin{equation}
T_{R}=(\frac{45}{2\pi^{2}g_{*}})^{\frac{1}{4}}(\Gamma_{\phi}M_{p})^{\frac{1}{2}}
\end{equation}          
where $M_{p}\approx2.4\times10^{18}$ Gev is the reduced Plank mass
[29] and $g_{*}$ is the effective number of relativistic degrees of
freedom at reheating temperature. For SM we have $g_{*}$= 106.75, and
for MSSM $g_{*}$= 228.75. If the inflaton dominantly couples to
$N_{i}$, the branching ratio of this dacay process is
taken as BR $\sim 1$, and the produced baryon asymmetry of the universe can be
calculated by the following relation [30],
\begin{equation}
Y_{B}=\frac{n_{B}}{s}= CY_{L}= C\frac{3}{2}\frac
{T_{R}}{M_{\phi}}\epsilon_{1}
\end{equation}

where $Y_{L}$ is the lepton asymmetry genarated by CP-violating
out-of-equilibrium decays of heavy neutrino $N_{1}$ and $T_{R}$ is the reheating
temperature. The fraction C has the value C=-28/79 for SM and C=-8/15 in
the MSSM. The observed baryon asymmetry measured in WMAP data,
$\eta_{B}=\frac{\eta_{B}}{\eta_{\gamma}}=6.5\times10^{-10}$ [3], 
where s=7.04 $n_{\gamma}$ is ralated to
$Y_{B}=n_{B}/s=8.6\times10^{-11}$ in eq.(29). From eq.(29) the
connection between $T_{B}$ and $M_{\phi}$ is expressed as,
\begin{equation}
T_{R}=(\frac{2Y_{B}}{3C\epsilon_{1}}) M_{\phi}
\end{equation}
The above expression is supplemented by two more boundary
conditions [28]: (i) lower bound on inflaton mass, $M_{\phi}>2M_{1}$
coming from allowed kinematics of inflaton decay to two right-handed Majorana
neutrinos $N_{1}$, and (ii) an upper bound for the reheating
temperature, $T_{R}\leq 0.01 M_{1}$ coming from out-of-thermal
equilibrium decay of $N_{1}$. Using the observed central value of the
baryon asymmetry $Y_{B}$ and theoretical prediction of CP asymmetry
$\epsilon_{1}$ in eq.(30), one can establish the relation
between $T_{R}$ and  $M_{\phi}$ for each neutrino mass model. The
lightest right-handed neutrino mass $M_{1}$ and the CP asymmetry $\epsilon_{1}$
neutrino mass models are used in the calculation of theoretical
bounds: $T^{min}_{R} < T_{R} \leq T^{max}_{R}$ and $M^{min}_{\phi} <
  M_{\phi}\leq M^{max}_{\phi}$ following eq.(19) along with other two
  boundary conditions cited above. Only those models which satisfy the
  constraints $T^{max}_{R} > T^{min}_{R}$ and $M^{min}_{\phi} <
  M^{max}_{\phi}$ simultneously can survive in the non-thermal
  leptogenesis.

%\newpage

%%%%%%%%%%%%%%%%%%%references%%%%%%%%%%%%%%%%%%%%%%%%%%%%%%%%%%%%%%%%%%%%%%


\begin{thebibliography}{20}

\bibitem{ref1} Shaun A. Thomas, Filipe B. Abdalla, Ofer Lahav,
  ``Upper Bound on 0.28 eV on Neutrino Masses from the Largest
  Photometric Redshift Survey,'' Phys Rev. Lett, Vol.{\bf 105}, 2010,
  pp. 031301; Jubien Lesgougues, ``Galaxies Weigh in on neutrinos,''
  Physics, Vol {\bf 3}, 2010, pp. 57-59.
\bibitem{ref2} S. Pascoli, S.T.  Petcov, ``Majorana Neutrinos, Neutrino
  Mass Spectrum and the  $<M>\sim0.001$ eV Frontier in Neutrinoless
  Double Beta Decay,''  Phys. Rev. D, Vol.{\bf 77} 2008, pp.113003, and
  further refernces there in.
\bibitem{ref3} D.N Spergel et.al., Astrophysics. J. Suppl. {\bf 148},
  175 (2003).
\bibitem{ref4} P.F. Harrison, W.G. Scott, 2002, ``$\mu-\tau$ Reflection
  Symmetry in Lepton Mixing and Neutrino Oscillation,'' Phys. Lett. B,
  {\bf 547}, pp. 219-228; Lam, C.S., 2005, ``Neutrino $\mu-\tau$
  symmetry  and Inverted Hierarchy,'' Phys. Rev. D., {\bf 71},
  pp. 093001; Grimus, W., Lavoura, L., 2007, ``A Three-Parameters
  Models for the Neutrino Mass Matrix,'' J. Phys. G {\bf 34(7)},
  pp. 1757-1769. 
\bibitem{ref5} G. Altarelli, F. Feruglio,  ``Neutrino Maases and Mixings:
  A Theoretical Perspective,''  Phys. Rep, Vol. {\bf 320}, 1999, pp. 295-325.
\bibitem{ref6}Z. Maki, M. Nakagawa, S. Sakata,  ``Remarks on the Unified
  Model of Elementary  Particles, '' Prog. Teor. Phys., Vol. {\bf 28}, 1962,
  pp. 87028; B. Pontecorvo, ``Neutrino Experiments and the Problems of
  Conservation of lepton Charge,''  Zh. Eksp. Teor. Fiz. Vol.{\bf 53},  1968,
  pp. 1717-1725 [Sov. Phys. JETP. Vol. {\bf 26}, 1968, pp. 984-988].
 \bibitem{ref7}P.F. Harrison, D.H. Perkins, W.G. Scott, ``Tri-Bimaximal
  Mixing and the Neutrino Oscillation Data,''
  Phys, Lett. B, Vol. {\bf 530}, 2002, pp. 167-178. P.F. Harrison,
  W.G. Scott, ``Permutation Symmetry, Tri-Bimaximal Neutrino Mixing
  and $S_{3}$,''  Phys. Lett. B,Vol. {\bf 557}, 2003, pp. 76-90.
\bibitem{ref8}B.T. Cleveland et al., Astrophysics. J. {\bf 496} (1998)
  505; J.N. Abdurashitov et al [SAGE Collaboration], J. Exp. Theor. Phys
 {\bf 95} (2002) 181 [astro-ph/0204245]; T. Kirsten et al [GALLEX and GNO
    Collaborations], Nucl Phys. B Suppl.) {\bf 118} (2003) 33;
  C. Cattadori, N. Ferri and L. Pandola, Nucl. Phys. B (Proc. Suppl.)
 {\bf 143} (2005) 3; T. Schetz, M. Tortola,
 J.W.F. Valle,  ``Three-Flavour Neutrino Oscillation Update,'' New
 J. Phys.  Vol. {\bf 10}, 2008,  pp. 113011.;  ``Global Neutrino 
  Data and Recent Reactor Fluxes: Status of Three-Flavour Oscillation
  Parameters,'' New J. Phys, Vol. {\bf 13}, 2011, pp. 063004.
M.C. Gonzales-Garcia, M. Maltoni, ''Phenomenology with
Massive neutrinos,''  Phys Rep. Vol. {\bf 460}. 2008;
   J.W.F. Valle.  ``Understanding and Probing Neutrino
   Oscillation.'' Invited Talk in Neutrino-2010, Athens, June
   14, 2010; A. Bandyopadhya et al., ``Physics at a Future Neutrino
   Factory and Super-Beam Facility,'' Rep. Prog. Phys. Vol. {\bf 72},
   2009, pp. 106201; S.T. Petcov, presented at  2011BCVSPIN, Hue,
   Vietnam, 27 July 2011,  ``Massive Neutrinis, Neutrinos
   Mixing, Oscillation, Leptonoc CP-Violation and Beyond.''
\bibitem{ref9} Ngouniba. Ki Francis and Ngangkham. Nimai
Singh, ``Quasi-Degenerate Neutrino Masses  with Normal and Inverted
Hierarchy,'' Journal of Modern Physics, {\bf Vol-2}, Number-11,
November 2011, pp 1280-1284.
\bibitem{ref10}N.Nimai Singh, Monisa Rajkhowa, Abhijit Borah,  ``Lowering
  Solar Mixing Angle in Inverted Hierarchy without charged Lepton
  Corrections,'' J. Phys. G: Nucl. Part. Phys. {\bf Vol.34}, 2007,
  pp. 345-351.;  ``Deviation From Tri- Bimaximal Mixing Through Flavour
  Twisters in Inverted and Normal Hierarchical Neutrino Mass Models,''
  Pramana J. Phys, {\bf Vol.69}, 2007, pp. 533-549.
\bibitem{ref11}Gell-Mann, M., Ramond, P., and Slansky, R., 1979,
  ``Supergravity, Proceeding of the Workshop,'' Stony Brook, New York,
  1979, Edited by Nieumenhuizen, P. Van, and
  Freedman, D.,(North-Holland, Amsterdam, 1979) Mahapatra, R.N.,and
  Senjanovic, G., 1980, ''Neutrino Mass and Spontaneous Parity
  Non-conservation,'' Phys. Rev. Lett., {\bf 44(14)}, pp. 912-915.


\bibitem{ref12} Asmaa Abada, Sacha Davidson, Francois-Xavier
  Josse-Michaux, Marta Losada and Antonio Riotto, ``Flavour Issues in
  Leptogenesis,'' arxiv: hep-ph/0601083v3, 5 Jan 2007.
\bibitem{ref 13} Enricho Nardi et al, ``The importance of flavour in
  Leptogenesis,''

\bibitem{ref14} Takeshi Fuguyama, Tatsuru Kikuchi and Toshiyuki Osaka,
  ``Non-thermal Leptogenesis and a prediction of Inflaton Mass in a
  Supersymmetric SO(10) Model,'' {\bf arxiv: hep-ph/053101v2 (2005)}.
\bibitem{ref15} Ng.K. Francis, N. Nimai Singh, H, Zeen Devi and Amal
  Kr Sarma, ``Constraints on the Validity of Neutrino Mass Models
  through Thermal and Non-thermal leptogenesis,'' IJAP {\bf Vol-1}
  Number-2. (November 2011) pp. 145-160.

\bibitem{ref16}Fukugita, M., and Yanagita, T., 1986, ``Baryogenesis
  without Grand Unification,'' Phys, Lett. B., {\bf 174(1)}, pp.45-47.
 
\bibitem{ref17}Luty, M.A. 1992, ``Baryogenesis via
  Leptogenesis,'' Phys. Rev. D, 1992, {\bf 45}, pp. 455-465.
\bibitem{ref18}Kolb, E.W.,Turner, M.S.,  The Early
  Universe. Addision-Wesely, New York.

\bibitem{ref19}Covi, L., Roulet, E., and Vissani, F., 1996,  ``CP Violating
  Decay in Leptogenesis Scenarios,'' Phys. Lett. B.,{\bf 384}, pp. 169-174;
  Pilaftis, A., 1997,  ``CP Violation and Baryogenesis due to Heavy
  Majorana Neutrinos,''  Phys. Rev. D., {\bf 56}, PP.54315451;
  Buchmuller, W., and Plumacher, M., 1998,
 ''CP Asymmetry in Baryogenesis Neutrino Decay,'' Phys. Lett. B., {\bf
  431}, pp. 354-362.
\bibitem{ref20} Davidson. S., Nardi. E.N.Y., 2008, ``Leptogenesis,''
  Phys. Report., {\bf 466}, pp. 105-177.
\bibitem{ref21} Pilaftis, A., 1997, ``CP Violation and Baryogenesis
  due to Heavy Majorana Neutrinos,'' Phys. Rev. D., {\bf 56},
  pp. 5431-5451; Pilaftis, A., Underwood, T.E.J., 2004, ``Resonant
  Leptogenesis,'' Nucl. Phys. B., {\bf 692}, pp. 303-345.
\bibitem{ref22} Khlebnikov, S.Y., Shaposhnikov, M.E., 1998, ``The
  Statistical Theory of Anomalous Fermion Number Non-conservation,''
  Nucl. Phys. B., {\bf 308}, pp. 885-912; Buchmuller, W., Pecci,
  R. D.m and Yanagida, T., 2007, ``Leptogeensis as the Origin of
  Matter,'' Ann. Rev. Nucl. Part. Sci. 55. pp 311-355.
\bibitem{ref23} Bari, P.D., 2005, ``Seesaw Geometry and
  leptogenesis,'' Nucl. Phys. B., {\bf 27} pp. 318-354; Buchmuller,
  W., Bari, P. D., and Plimacher, M., 2003, ``The Neutrino Mass Window
  for Baryogenesis,'' Nucl. Phys. B., {\bf 665}, pp. 445-468.
\bibitem{ref24} Branco, G. C., Felipe, R. G., Joaquim, F.R and Rebelo,
M.N., 2002, ``Leptogenesis, CP Violation and Neutrino Data: What can
we learn,'' Nucl. Phys. B., {\bf 640}:pp. 202-232; Akhmedov, E. K.,
Frigerio, M., and Smirnov, A.Y., 2003, JHEP, 0309, pp. 021-051;
Adhikary, B., and Ghosal, A., 2008, ``Neutrino U($\epsilon_{3})$, Cp
Violation and leptogenesis in a seesaw Typw Softly Broken $A({4}$
Symmetry Model,'' Phys. Rev. D., {\bf 78}, pp. 073007; Buccella, F.,
Falcone, D., and Oliver, L., 2008, ``Leptogenesis within a Generalized
Quark-Lepton Symmetry,'' Phys. Rev. D., {\bf 77}, pp. 033002.
\bibitem{ref25} Babu, K. S., Bachri, A., and Aissaoui, H, 2006.,

  ``Leptogenesis in minimal left-right symmetric models,''
  Nucl. Phys. B., {\bf 738}, pp. 76-92.
\bibitem{ref26}Adaba, A., Aissaoui, H., and Losada, M., 2005, ``A Model of
  leptogenesis at the TeV scale,'' Nucl.Phys. B., {\bf 728}
  pp. 55-66.; Vives, O., 2006. ``FlavouedLeptogenesis: a successful
  Thermal Leptogenesis with $N_{1}$ Mass below $10^{8}$ GeV,''
  Phys.Rev. D., {\bf 73} pp. 073006; Abadam A., Davidson, S., Ibarra,
  A., Josse-Michaux, F. X., Losada. M., Riotto, A., 2007, ``Flavour
  matters in Leptogenesis,'' JCAP, 0604, pp. 004-040; Nardi, E., Nir,
  Y., Roulet, E., and Racker, J., 2006, ``The Important of Flavour in
  leptogenesis,'' JHEP, 0601, pp. 164-171.
\bibitem{ref27}Lazarides, G., and Shafi, Q., 1991, ``Origin of Matter
  in the Inflationary Cosmology,'' Phys. Lett. B., {\bf 258}
  pp. 305309; Kumekawa, K., Moroi, T., Yanagida, T., 1994, ``Flat
  Potential for Inflaton with a Discrete ?R? Invariance in
  Supergravity,'' Prog. Theor. Phys., {\bf 92}, pp. 437-448; Guidice,
  G. F., Peloso, M., Riotto, A., 1999; ``Production of Massive
  Fermions at Preheating and Leptogenesis,'' JHEP., {\bf 9908},
  pp. 014-043; Asaka, T., Hamaguchi, K., Kawasaki, M., and Yanagida,
  T, 1999, ``Leptogenesis in Inflaton Decay,'' Phys. Rev B., {\bf 464}
  pp. 12-18; ``Leptogenesis in Inflationary Universe,'' Phys. Rev. D.,
  2000, {\bf 61(8)} pp. 083512; Aska, T., Nielsen, H. B., and
  Takanishi, Y., 2002, ``Non-Thermal Leptogenesis from the Heavier
  Majorana Neutrinos,'' Nucl. Phys. B., {\bf 647}, pp. 252-274;
  Mazumdarm A., 2004, CMB Constraints on Non-Thermal Leptogenesis,''
  Phys. Lett. B., {\bf 580}, pp. 7-16; Fukuyama, T., Kikuchi, T., and
  Osaka, T., 2005, ``Non-Thermal Leptogenesis and a Prediction of
  Inflaton Mass in a Supersymmetric SO(10) Model,'' JCAP, 0506,
  pp. 005-014.
\bibitem{ref28}  Nanotopolous, G., 2006, ``Non-Thermal and Baryon
  Asymmetry in Different Neutrino Mass Models,'' Phys. Lett. B., {\bf
    643}, pp. 279283.
\bibitem{ref29} Steffen, F.D., 2008, `` Probing the relating
  Temperature ar Colliders and with Primodial Nucleosynthesis,''
  Phys. Lett. B., {\bf 669} pp. 74-80.
\bibitem{ref30} Buchmuller, W., Peccei, R.D., and Yanagida, T., 2005,
  ``Leptogenesis as Origin of Matter,''
  Annual. Rev. Nucl. Part. Sci. {\bf 55}, pp. 311-355.

\end{thebibliography}
\end{document}